\documentclass[acmsmall,screen,nonacm]{acmart}

\usepackage{amsmath,amsfonts,amsthm}
\usepackage{algorithm}
\usepackage{algpseudocode}
\usepackage{array}
\usepackage{textcomp}
\usepackage{xcolor}
\usepackage{balance}
\usepackage{xspace}
\usepackage{enumitem}
\usepackage{hyperref}
\usepackage[many]{tcolorbox}
\usepackage{subfigure}
\usepackage{booktabs}
\usepackage[normalem]{ulem}
\usepackage{siunitx}
\usepackage{colortbl}
\usepackage{listings}
\usepackage{graphicx}
\usepackage{wrapfig}
\usepackage{multicol}
\usepackage{multirow}
\usepackage{threeparttable}

\usepackage[normalem]{ulem} %

\definecolor{xrclientcolor}{RGB}{128, 222, 246}
\definecolor{networkcolor}{RGB}{206, 155, 219}
\definecolor{statecolor}{RGB}{164, 210, 173}
\definecolor{objectcolor}{RGB}{228, 228, 146}
\definecolor{avatarcolor}{RGB}{237, 137, 177}
\definecolor{qoecolor}{RGB}{191, 169, 205}

\newcommand{\xrclients}[1]{\textcolor{xrclientcolor}{\textbf{#1}}}
\newcommand{\networking}[1]{\textcolor{networkcolor}{\textbf{#1}}}
\newcommand{\statesync}[1]{\textcolor{statecolor}{\textbf{#1}}}
\newcommand{\sharedobject}[1]{\textcolor{objectcolor}{\textbf{#1}}}
\newcommand{\avatar}[1]{\textcolor{avatarcolor}{\textbf{#1}}}
\newcommand{\qoe}[1]{\textcolor{qoecolor}{\textbf{#1}}}

\def\BibTeX{{\rm B\kern-.05em{\sc i\kern-.025em b}\kern-.08em
    T\kern-.1667em\lower.7ex\hbox{E}\kern-.125emX}}
\begin{document}

\title{When Shared Worlds Break: Demystifying Defects in Multi-User Extended Reality Software Systems}

\author{Shuqing Li}
\orcid{0000-0001-6323-1402}
\affiliation{%
  \institution{The Chinese University of Hong Kong}
  \city{Hong Kong}
  \country{China}
}
\email{sqli21@cse.cuhk.edu.hk}

\author{Chenran Zhang}
\affiliation{%
  \institution{Harbin Institute of Technology}
  \city{Shenzhen}
  \country{China}
}
\email{220110514@stu.hit.edu.cn}

\author{Binchang Li}
\orcid{0009-0008-5995-4040}
\affiliation{%
  \institution{Harbin Institute of Technology}
  \city{Shenzhen}
  \country{China}
}
\email{24s151125@stu.hit.edu.cn}

\author{Cuiyun Gao}
\orcid{0000-0003-4774-2434}
\affiliation{%
  \institution{Harbin Institute of Technology}
  \city{Shenzhen}
  \country{China}
}
\email{gaocuiyun@hit.edu.cn}

\author{Michael R. Lyu}
\orcid{0000-0002-3666-5798}
\affiliation{%
  \institution{The Chinese University of Hong Kong}
  \city{Hong Kong}
  \country{China}
}
\email{lyu@cse.cuhk.edu.hk}

\begin{abstract}
Multi-user Extended Reality (XR) systems enable transformative shared experiences but introduce unique software defects that compromise user experience.
Understanding software defects in multi-user XR systems is crucial for enhancing system reliability, yet remains underexplored. 
To fill the gap,
this paper presents the first large-scale empirical study of multi-user XR defects, analyzing 2,649 real-world bug reports from diverse sources, including developer forums, GitHub repositories, and app reviews on mainstream XR app stores. 
Through rigorous qualitative analysis using iterative open coding, we develop a comprehensive taxonomy that classifies multi-user XR bugs along three dimensions: Symptom Manifestation, Root Cause Origin, and Consequence Severity. 
Our findings reveal that synchronization inconsistencies and avatar-related anomalies are the most prevalent symptoms, while network/synchronization logic defects and session management flaws emerge as dominant root causes. 
Critically, over 34\% of analyzed bugs lead to severe consequences that fundamentally break the shared experience, including system crashes, persistent disconnections, and complete interaction breakdowns, etc. 
We also identify concerning privacy and health implications unique to multi-user XR contexts. 
Based on our findings of defect analysis, we provide actionable recommendations for developers, platform vendors, and researchers. 
Our results demonstrate that multi-user XR systems face distinct challenges at the intersection of distributed systems, real-time 3D interaction, and immersive experiences, necessitating specialized approaches to testing, debugging, and quality assurance.
\end{abstract}
\maketitle

\section{Introduction}
\label{sec:introduction}

Extended Reality (XR) technologies, encompassing Virtual Reality (VR), Augmented Reality (AR), and Mixed Reality (MR)~\cite{csur-xr-tertiary-review}, are experiencing unprecedented growth across diverse application domains, from collaborative workspaces~\cite{lee2021xr} and social platforms~\cite{doukianou2024framework} to education~\cite{vasilchenko2020collaborative} and remote medical operation collaboration~\cite{burian2023using}. 
The transformative potential of multi-user XR lies in its ability to create shared, embodied experiences within persistent virtual worlds, fundamentally distinguishing it from single-user XR applications or traditional online collaboration tools~\cite{nguyen2020user}. 
In these shared environments, multiple users can simultaneously interact, communicate, and manipulate virtual objects, creating new paradigms for remote collaboration and social interaction.

However, the very characteristics that make multi-user XR compelling, including \textit{real-time synchronization}, \textit{distributed state management}, \textit{embodied presence through avatars}, and \textit{low-latency interaction requirements}, also introduce a complex landscape of potential software defects. 
These bugs not only inconvenience users but can lead to severe failures in collaborative tasks, significant economic losses in commercial applications, and ultimately, user abandonment of XR platforms~\cite{software-testing-for-ar-vr-ensuring-bug-free-experiences}. 
The impact of bugs is amplified in multi-user scenarios where a single user's issue can cascade through the shared experience, disrupting all participants. 
For instance, when one user cannot join a session or experiences avatar rendering anomalies, it affects not just their individual experience but the collective presence and interaction dynamics of the entire group.

The unique challenges of multi-user XR systems stem from their inherent complexity. 
Maintaining state consistency across geographically distributed users under varying network conditions, managing concurrent access to shared resources, ensuring accurate and responsive avatar interactions, and handling the heterogeneity of hardware and software configurations create numerous opportunities for failures that are either unique to or significantly exacerbated by the multi-user context. 
These challenges go beyond traditional distributed systems or multiplayer games. 
The immersive nature of XR means that synchronization errors, latency spikes, or rendering inconsistencies can cause not just functional failures but also physical discomfort~\cite{biswas2024you} and break the sense of shared presence that is fundamental to the XR experience.

Despite extensive research on software bugs in various domains, including general open-source software~\cite{li2006have}, mobile applications~\cite{xiong2023empirical}, web systems~\cite{ocariza2013empirical}, and even aspects of single-user XR~\cite{li2020exploratory}, there remains a significant gap in our systematic understanding of bugs specifically arising from multi-user interactions in XR environments. 
This knowledge deficit hinders developers from receiving targeted guidance for bug prevention and mitigation, prevents platform vendors from adequately supporting robust multi-user functionality, and limits researchers from developing effective automated quality assurance tools. 
The fragmented nature of bug reporting across diverse platforms, i.e., from GitHub repositories to engine-specific forums and device community pages, further complicates the aggregation of knowledge and identification of common patterns.

To address this gap, we present the first large-scale empirical study of multi-user related bugs in XR software systems. Our investigation is guided by the following research questions:

\begin{description}
\vspace{-0.25em}
\item[\textbf{RQ1:}] What are the common observable symptoms of multi-user related bugs in XR software systems?
\item[\textbf{RQ2:}] What are the underlying root causes contributing to these multi-user XR bugs?
\item[\textbf{RQ3:}] What are the typical consequences of these bugs on user experience and system functionality?
\item[\textbf{RQ4:}] What are the specific privacy and health implications associated with multi-user XR bugs?
\vspace{-0.25em}
\end{description}

To answer these questions, we conduct a comprehensive empirical analysis of bug reports collected from diverse sources including developer forums (Unreal Engine, Unity, VIVE), user communities (Stack Overflow, Meta, Steam), application store reviews (Oculus, Steam), and GitHub issue tracking systems. 
Through rigorous qualitative analysis involving multiple rounds of iterative open coding~\cite{book:open-coding-16}, we develop comprehensive taxonomies that characterize multi-user XR bugs along three key dimensions: symptom manifestation, root cause origin, and consequence severity.

Our analysis reveals that synchronization problems, interaction failures, and avatar representation errors constitute the most prevalent symptom categories, covering 34.8\%(922/2,649) analized bugs, with network instability, synchronization logic defects, and permission control flaws emerging as dominant root causes, covering 47.2\%(50/106) analized bugs. 
These findings highlight that multi-user XR is not merely an extension of single-user XR or traditional multiplayer gaming, it represents a distinct paradigm with unique bug characteristics. 
For example, incorrect avatar rendering in a shared space not only represents a technical failure but fundamentally breaks social presence, a phenomenon not central to traditional distributed systems. 
The synchronization of fine-grained embodied interactions, such as collaborative object manipulation using hand tracking, introduces complexities beyond typical app/game state synchronization.

In summary, the contributions of this paper are:

\begin{enumerate}[leftmargin=*]
\item 
We present the first large-scale, systematic study of bugs arising from multi-user XR software systems, based on analysis of real-world bug reports from multiple sources.

\item 
We develop and present two comprehensive taxonomies detailing bug symptoms and root causes, quantified by their prevalence in our dataset, providing a structured framework for understanding multi-user XR defects.
We provide detailed analysis of how these bugs affect user experience, system functionality, and overall application viability, including 
privacy and health-related concerns unique to multi-user XR.

\item 
We derive practical recommendations for developers (design patterns, testing focuses), platform vendors (API improvements, better default behaviors), and researchers (future research directions, tool development opportunities).
\end{enumerate}

\begin{figure}[t!] 
 	\centering 
 	\includegraphics[width=0.7\columnwidth]{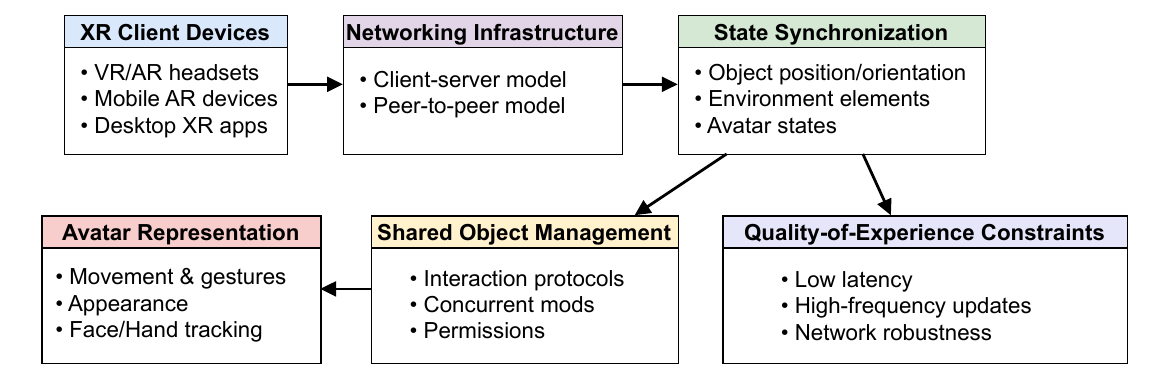} 
        \vspace{-1.5em}
 	\caption{The architecture of multi-user XR software systems}
 	\label{fig:multi-user-xr-arch}
       \vspace{-1.5em} 
 \end{figure}

\section{Preliminaries on Multi-User XR}

Extended Reality (XR) encompasses Virtual Reality (VR), Augmented Reality (AR), and Mixed Reality (MR) technologies that have evolved from single-user experiences to collaborative platforms enabling multiple users to interact within shared digital environments~\cite{csur-xr-tertiary-review, etelligens2025}. These multi-user XR systems facilitate applications across social interaction, education, training, and gaming domains~\cite{lee2021xr, doukianou2024framework, vasilchenko2020collaborative}.
As demonstrated in Figure~\ref{fig:multi-user-xr-arch}, multi-user XR architectures typically comprise: (1) \xrclients{XR Client Devices}, the terminals users employ to access the shared experience, such as VR/AR headsets, mobile AR devices, and desktop XR applications; (2) \networking{Networking Infrastructure} using client-server or peer-to-peer models via frameworks like Unity Netcode~\cite{unity-netcode}, Photon~\cite{photonengine}, or WebRTC; (3) \statesync{State Synchronization} mechanisms, which act as a central hub for managing and distributing the positions, orientations, and properties of all shared objects and avatars; (4) \sharedobject{Shared Object Management} protocols, which govern how users interact with virtual elements, including interaction protocols, concurrent modifications, and permissions; (5) \avatar{Avatar Representation} systems, responsible for synchronizing user avatars, including their movements, gestures, appearance, and even face/hand tracking data, with these representations often being influenced by interactions with shared objects; and (6) \qoe{Quality-of-Experience Constraints}, a set of overarching performance requirements such as low latency, high-frequency updates, and network robustness, which impose strict challenges on the entire system design.
The technical challenges are distinct from traditional distributed systems. Ultra-low latency (sub-20ms) is critical to prevent motion sickness~\cite{biswas2024you}. Systems must handle high-frequency 6DOF tracking updates while maintaining synchronization across heterogeneous devices~\cite{tumler2022multi}. Critically, inconsistencies in the shared experience can break presence and cause user disorientation~\cite{stanney2020identifying}.

\section{Methodology}

This section describes our systematic approach to investigating multi-user bugs in XR systems. We employ a mixed-methods research design combining qualitative and quantitative analyses to comprehensively understand the nature, causes, and impacts of these bugs.

\subsection{Research Questions}

As discussed in Section~\ref{sec:introduction}, our empirical study is guided by four research questions,
moving from observable phenomena (RQ1) to underlying technical issues (RQ2), then to impacts on users and systems (RQ3), and finally to critical concerns unique to XR environments (RQ4).

\subsection{Bug Report Collection}

\subsubsection{Source Selection and Rationale}

As shown in Table~\ref{tab:data_sources}, we adopt a comprehensive multi-source data collection strategy to capture the diversity of multi-user XR bugs across different platforms, engines, and user contexts. Our data sources are strategically selected to represent different stakeholder perspectives:
GitHub issue trackers, developer forums, user communities, and reviews on app stores.

\subsubsection{Data Extraction Process}

The data extraction followed a systematic protocol:

    \textbf{Initial Search and Retrieval:} We search with keywords ``XR'', ``extended reality'', ``VR'', ``virtual reality'', ``AR'', ``augmented reality'', ``MR'', ``mixed reality'', ``multi'', ``multiplayer'', ``multi-user'', ``bug'', and their synonyms.
    \textbf{Relevance Filtering:} Each retrieved item is assessed for relevance based on two criteria: (a) the issue must be related to XR applications, and (b) the bug must manifest specifically in multi-user contexts or be exacerbated by multi-user interactions.
    Bug reports collected from \textbf{most of the platforms} (except app reviews) are technically deep, thus two analyzers analyze their relevance manually. Both analyzers have more than three years of programming experience and sufficient knowledge on multi-user XR.
    Given that \textbf{app reviews} are numerous but typically lack technical depth, we employ an iterative machine learning approach for filtering. Initial keyword filtering is followed by multiple rounds of LLM-based relevance classification using DeepSeek v3, with manual validation of a 10\% sample from each round to prevent false negatives. The process was terminated when the false positive rate exceeded acceptable thresholds (0.2\%).

\subsubsection{Dataset Characteristics}

Our final dataset comprises 2,649 unique bug reports after removing duplicates and irrelevant entries. Table \ref{tab:data_sources} summarizes the distribution of bug reports across sources.
The dataset covers major XR ecosystems (Oculus/Meta Quest, SteamVR, VIVE, Pico, etc.), development engines (Unity, Unreal Engine, etc.), and frameworks (networked-aframe, etc.). Bug reports span from 2018 to 2025.

\begin{table}[t]
\centering
\caption{Distribution of Bug Reports by Source}
\label{tab:data_sources}
\begin{threeparttable}
\resizebox{0.6\linewidth}{!}{
\begin{tabular}{l|crr}
\toprule
\textbf{Source Category} & \textbf{Platform} & \textbf{Searched} & \textbf{Filtered} \\
\midrule
GitHub Issues & GitHub & 11,955 & 109 \\
\midrule
Developer Forums & Unreal Engine & 9,165 & 6 \\
  & Unity & 1,676 & 5 \\
  & Meta Community Forums & 4,594 & 48 \\
  & Others\tnote{a} & 373 & 6 \\
\addlinespace
\cline{2-4}
\addlinespace
  & Total & 15,908 & 65 \\
\midrule
User Communities & Reddit & 264 & 30 \\
  & Steam & 20,575 & 29 \\
  & Others\tnote{b} & 1,050 & 3 \\
\addlinespace
\cline{2-4}
\addlinespace
  & Total & 21,889 & 62 \\
\midrule
App Reviews & oculus-quest & 4,054 & 1,739 \\
  & oculus-rift & 246 & 111 \\
  & sidequest & 178 & 54 \\
  & steam-vr & 1,657 & 487 \\
  & Others\tnote{c} & 56 & 22 \\
\addlinespace
\cline{2-4}
\addlinespace
 & Total & 6,191 & 2,413 \\
\midrule
Total & --- & 55,943 & 2,649 \\
\bottomrule
\end{tabular}
}
\begin{tablenotes}
\footnotesize
\item[a] VIVE Forum, Apple Developer Forum
\item[b] Stack Overflow, Discord
\item[c] Google Play, Apple App Store, Viveport
\end{tablenotes}
\end{threeparttable}
\end{table}

\subsection{Data Analysis}

We employ a rigorous qualitative analysis approach grounded in empirical software engineering methodologies, complemented by quantitative analysis for pattern identification.

\subsubsection{Qualitative Analysis - Taxonomy Development}

We follow an iterative open coding process \cite{book:open-coding-16} involving three researchers with more than three years of software engineering experience and sufficient expertise in XR development.
Firstly, the three researchers analyze and code all bug reports independently. Codes were assigned for symptoms, root causes, consequences, and contextual factors.
Then the research team meets to compare initial codes, resolve discrepancies through discussion (i.e., refine categories, merging similar codes, splitting overly broad categories, and adding new codes for emerging patterns), and develop a preliminary coding scheme. 
After that, the team iterates to reanalyze all bug reports and then come back to discussions and cross-validation.
This process continued until saturation is reached on all bug reports, costing us four iterations.
Final codes were organized into a hierarchical taxonomy along three dimensions: Symptom Manifestation, Root Cause Origin, and Consequence Severity. Each category was defined with clear inclusion criteria and exemplar cases, further detailed in Section~\ref{sec:multiuser-empirical-results}.

\subsubsection{Quantitative Analysis}

Following taxonomy development, we conduct quantitative analyses to identify patterns and trends:

(1) \textbf{Frequency analysis}: We calculated the distribution of bugs across taxonomy categories to identify the most prevalent symptoms, root causes, and consequences.
(2) \textbf{Cross-tabulation analysis}: We examined relationships between symptoms and root causes to identify common bug patterns and diagnostic indicators.
(3) \textbf{Platform-specific analysis}: Where sample sizes permitted, we analyzed whether certain bug types were more prevalent on specific platforms or engines.
(4) \textbf{Severity distribution}: We assessed the proportion of bugs causing critical failures versus minor annoyances to understand the overall impact on user experience.

\subsubsection{Analysis of Privacy and Health Implications}

For RQ4, we employed a targeted analysis approach:
(1) \textbf{Keyword-based identification}: Bug reports are first filtered using privacy-related keywords (e.g., ``security'', ``privacy'', ``permission'', ``data leak'') and health-related keywords (e.g., ``health'', ``safety'', ``nausea'', ``discomfort'', ``motion sickness'').
(2) \textbf{Manual classification}: Each identified report is manually reviewed to confirm privacy or health relevance and classify the specific type of concern.
(3) \textbf{Impact assessment}: We manually analyze the potential severity and scope of privacy violations and health impacts, considering factors such as data sensitivity and physical discomfort levels.

\section{Results \& Analysis}
\label{sec:multiuser-empirical-results}

\subsection{RQ1: Symptoms of Multi-User XR Bugs}

\begin{table}[t]
\centering
\caption{Symptoms Categories and Statistics\tnote{1}}
\label{tab:symptom-statistics}
\vspace{-1em}
\begin{threeparttable}
\resizebox{0.6\columnwidth}{!}{
\begin{tabular}{l|crr}
\hline
\textbf{Main Categories} & \textbf{Subcategories} & \textbf{Count} & \textbf{Percentage} \\
\hline
State-Sync & Object Sync & 114 & 3.53\% \\
  & Avatar Sync & 131 & 4.06\% \\
  & Event-World Sync & 233 & 7.22\% \\
\hline
Interact-Control & Asymmetric & 60 & 1.86\% \\
  & Control Loss & 152 & 4.71\% \\
  & Unintended \& Mismatch & 82 & 2.54\% \\
\hline
Comm-Awareness & Audio & 223 & 6.91\% \\
  & Avatar Display & 150 & 4.65\% \\
  & Status Info & 55 & 1.70\% \\
\hline
Perf-Stability & System Crash & 338 & 10.47\% \\
  & Lag-Frame \& Freeze & 398 & 12.33\% \\
  & Connection & 1,081 & 33.50\% \\
\hline
Others & --- & 210 & 6.51\% \\
\hline
Total Reports & --- & 3,227 & 100\% \\
\hline
\end{tabular}
}
\begin{tablenotes}
\footnotesize
\item[1] \parbox[t]{0.5\columnwidth}{The user review data tend to lack technical details; we only use them in qualitative analysis, not in quantitative analysis.\strut}
\end{tablenotes}
\end{threeparttable}
\vspace{-1em}
\end{table}

Understanding the observable symptoms of multi-user bugs in XR systems is crucial for both practitioners and researchers, as these manifestations represent the first indicators of underlying quality issues that developers and users encounter. Through our systematic analysis of 2,649 bug reports, we identified a diverse spectrum of symptoms that uniquely characterize the failure modes of shared XR experiences. These symptoms not only affect functional correctness but also fundamentally compromise the sense of presence and shared reality that underpins collaborative XR applications.

\subsubsection{Overview of Symptom Distribution}

Table~\ref{tab:symptom-statistics} presents the distribution of observable symptoms across the four main categories identified in our taxonomy. State \& Synchronization Inconsistency (S1) is the most representative symptom category, accounting for 18.0\% (478/2,649) of the analyzed bugs. This is followed by Interaction \& Control Discrepancy (S2) at 11.1\% (294/2,649), Performance \& Stability Degradation (S4) at 16.2\% (428/2,649), and Communication \& Awareness Impairment (S3) at 68.6\% (1,817/2,649).

\subsubsection{State \& Synchronization Inconsistency (S1)}

\begin{figure}[t!] 
 \vspace{-0.5em} 
 	\centering
 	\subfigure[\parbox{0.8\linewidth}{Example of Object State Desynchronization}]{ 
 		\label{fig:ADP259911-object-state-desync} 
         \vspace{-1.5em} 
        \includegraphics[width=0.32\linewidth]{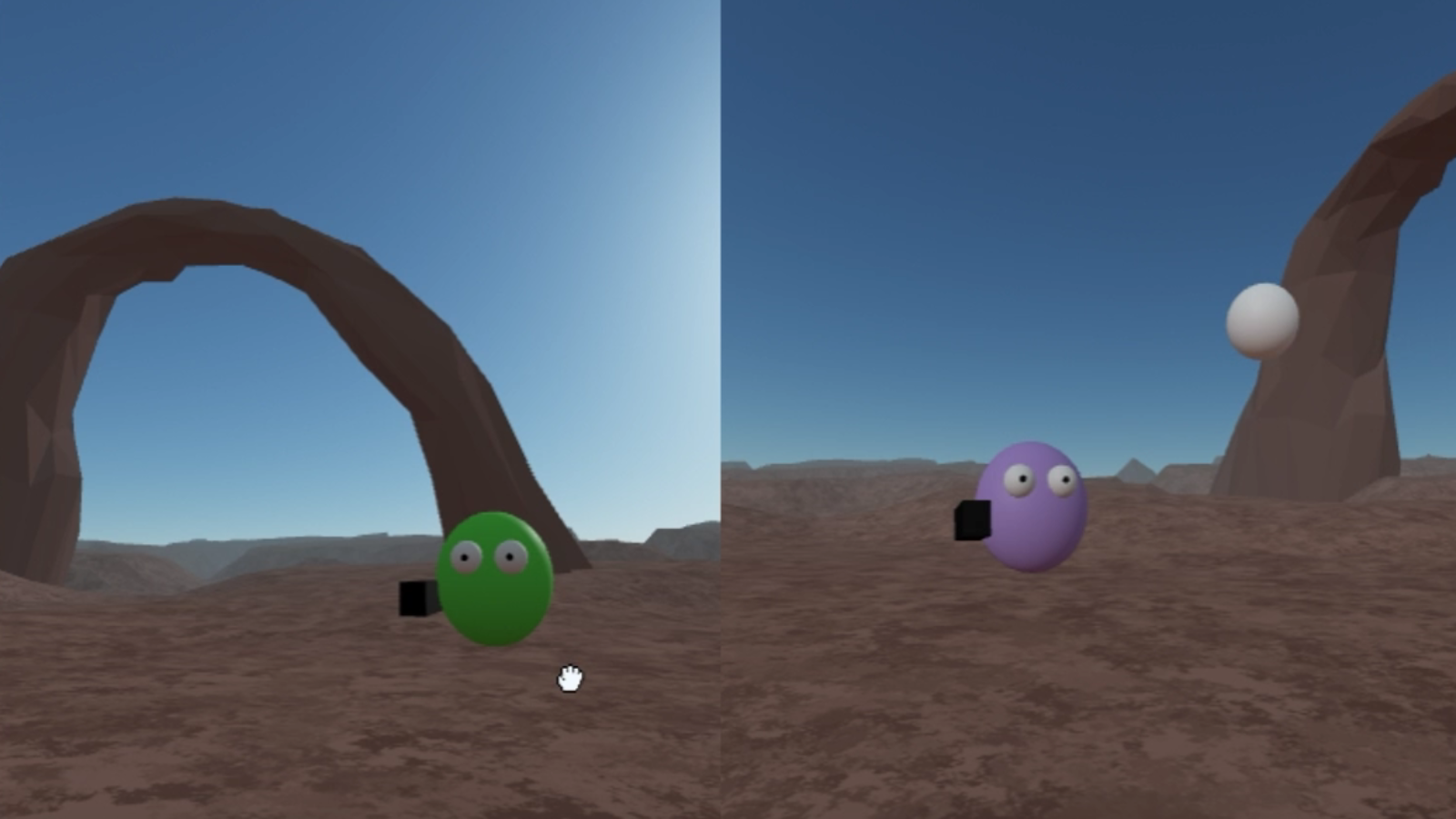}} 
 	\subfigure[\parbox{0.8\linewidth}{Example of Event Desynchronization}]{ 
 		\label{fig:event-example} 
 		\includegraphics[width=0.32\linewidth]{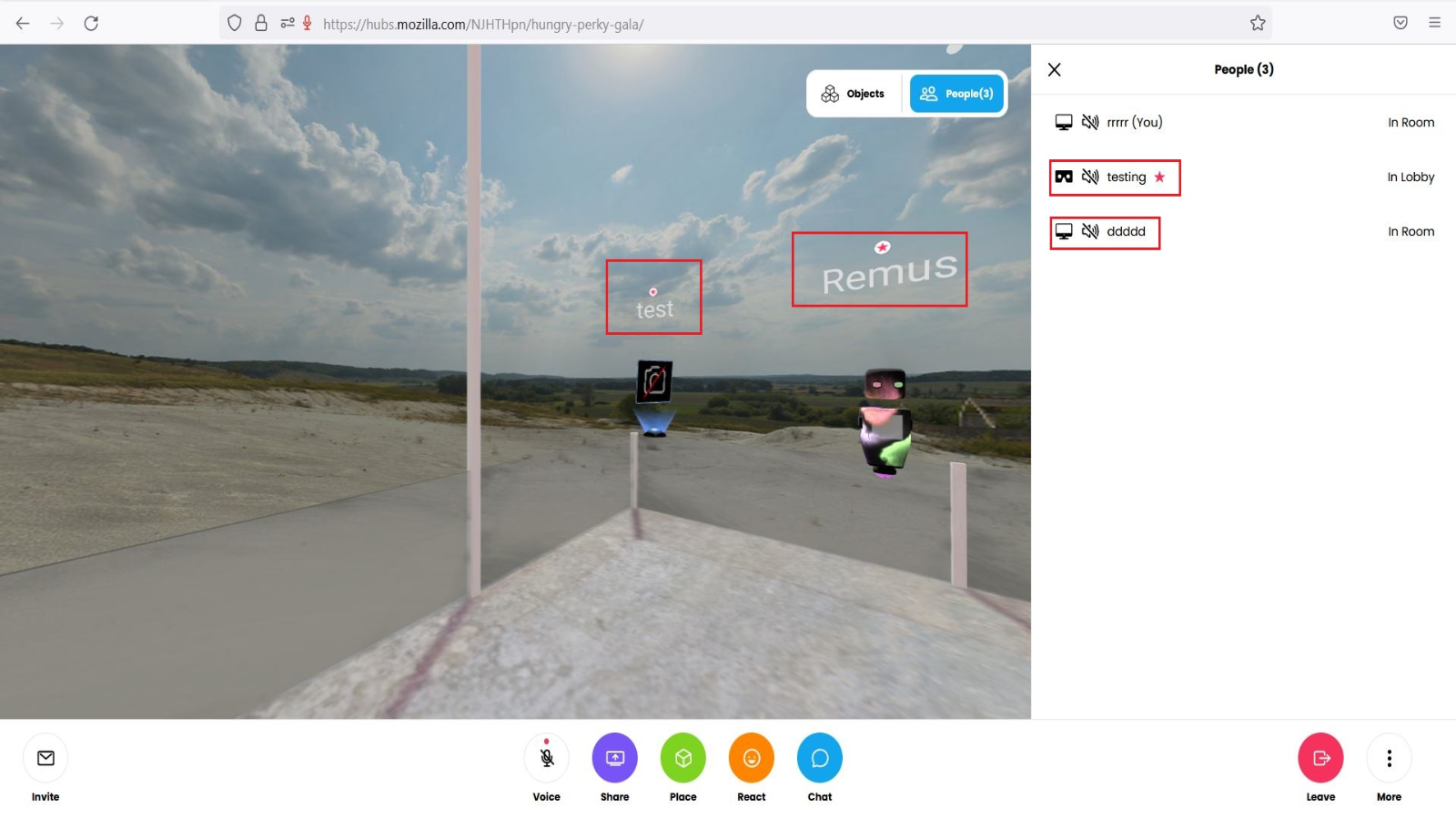}} 
 	\subfigure[\parbox{0.8\linewidth}{Example of Incorrect ``In Lobby'' Status}]{ 
 		\label{fig:Status-Info-example} 
 		\includegraphics[width=0.32\linewidth]{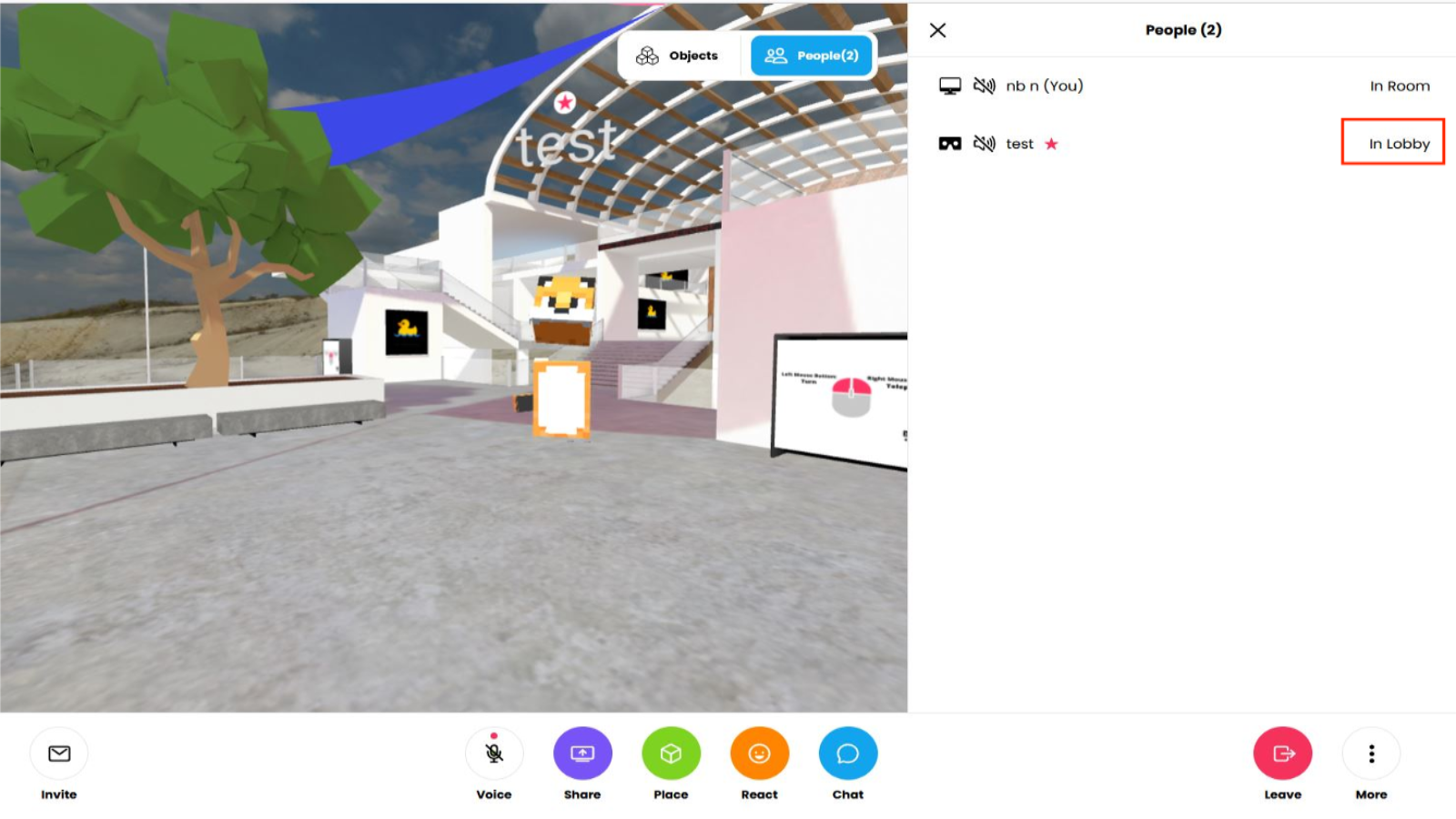}} 
      \vspace{-1.3em} 
       \caption{Example of State \& Synchronization Inconsistency and Interaction \& Control Discrepancy} 
      \vspace{-1.5em} 
 	\label{figs:xr-device} 
 \end{figure}

The state and synchronization issues underscores a fundamental challenge in multi-user XR: maintaining a coherent shared reality across distributed participants. Within this category, we observed four distinct sub-patterns:

\textbf{Object State Desynchronization (S1.1 Object Sync)} represents the most significant manifestation (114/478 bugs in S1). For example, sometimes ``grabbable objects owned by the master client cannot be moved by other clients,'' creating an asymmetric reality where actions performed by non-master clients fail to propagate~\cite{SII187773-object-state-desync}. 
Similarly, as shown in Figure~\ref{fig:ADP259911-object-state-desync}, bullets are invisible to the user who initializes them (left) but visible to everyone else (right)~\cite{ADP259911-object-state-desync}.
The prevalence of this symptom suggests that current state synchronization mechanisms struggle with the complex ownership models and dynamic state changes inherent in interactive XR environments.

\textbf{Avatar State Desynchronization (S1.2 Avatar Sync)} occurred in 131/478 S1 bugs, with manifestations ranging from incorrect spatial representation to broken embodiment.
Some users report that ``some avatars will display with the incorrect height to other players,'' a particularly disorienting issue in XR where spatial relationships directly impact social presence and interaction affordances~\cite{WZE463732-avatar-state-desync}. 
More severe cases, describe avatars becoming ``stuck in T-pose'', i.e., the default arms-out, T-shaped stance for rigging humanoid 3D models~\cite{BBZ270278-avatar-state-desync-2}.
This badly breaks the embodiment illusion. 
The high frequency of avatar-related symptoms reflects the critical role of embodied presence in XR and the technical complexity of synchronizing full-body tracking data, facial expressions, and gesture animations across network boundaries.

\textbf{Event Desynchronization (S1.3)} and \textbf{Inconsistent World View (S1.4) (Event-World Sync)} together accounted for the remaining 233/478 S1 bugs. 
Some developers report event desynchronization where (1) component deletions by a user are not synchronized across other users' views ~\cite{CEU902107-event-desync}, (2) updates to the display name are not reflected in the name tags~\cite{event-desync-3}, etc., leading to divergent world states. We can see an example of this desynchronization in Figure ~\ref{fig:event-example}. 
Another multi-user XR app demonstrates an inconsistent world view where ``procedurally generated levels result in different layouts for different users,'' fracturing the shared experience~\cite{event-desync-2}. 
These symptoms highlight the challenges of ensuring both discrete events and continuous world generation algorithms produce consistent results across all participants.

\subsubsection{Interaction \& Control Discrepancy (S2)}

The second category reveals fundamental issues in user interactions within shared XR spaces. Unlike traditional multi-user applications, XR systems must coordinate complex spatial interactions involving hand tracking, gaze, and full-body movements.

\textbf{Asymmetric Interaction Capabilities (S2.1 Asymmetric)} affect 60/294 S2 bugs. It manifests when users inexplicably have different abilities to interact with the shared environment. 
Developers report that ``PC users' movements are visible, but Oculus Go users' movements are not,'' creating a particularly frustrating asymmetry where cross-platform participants exist in different interaction hierarchies~\cite{UXV944060-asymmetric}. 
This symptom often correlates with platform-specific implementations that fail to consider diversed capability across XR devices.

\textbf{Loss/Hijack of Control (S2.2 Control Loss)} represents a severe symptom affecting 152/294 S2 bugs. A bug describes a scenario where ``a user loses the ability to move or rotate after another user joins,'' effectively trapping them in the virtual space~\cite{CTC541318-phy-safe}. Even more concerning is another one, where ``a joining client takes over the host's camera view,'' violating fundamental assumptions about user agency~\cite{DLU106275-SDK-API}. These symptoms are particularly jarring in XR, where loss of control can induce discomfort or even motion sickness.

\textbf{Unintended Interactions/Interference (S2.3)} and \textbf{Action/Feedback Mismatch (S2.4) (Unintended \& Mismatch)} round out this category. The Stack Overflow issue \#76449733 exemplifies interference where ``grabbing an object simultaneously results in both players believing they hold it,'' creating logical paradoxes in the shared space. The Beat Saber multiplayer bug demonstrates feedback mismatch where ``hits are not registered in multiplayer mode,'' fundamentally breaking the core gameplay loop.

\subsubsection{Communication \& Awareness Impairment (S3)}

While less frequent than state and interaction issues, communication impairments have disproportionate impact on the collaborative aspects of multi-user XR. This category encompasses the social and awareness mechanisms that enable meaningful shared experiences.

\textbf{Audio Malfunction (S3.1 Audio)} dominated this category (223/428 S3 bugs), reflecting the critical role of spatial audio in XR communication. One of bugs reports that ``only the first user in a room can transmit audio,'' effectively silencing all subsequent participants~\cite{RPQ265395-major-func}. The Meta Community Forums bug describing one-way audio transmission creates particularly confusing scenarios where users can hear but not be heard, breaking the reciprocal nature of conversation. 
While visual attention in XR is limited by field of view, audio becomes even more critical for maintaining awareness of other participants.

\textbf{Avatar Invisibility/Misrepresentation (S3.2 Avatar Display)} affected 150/428 S3 bugs. A bug report of completely invisible avatars represents a total breakdown of visual presence, while the VRChat bug describing avatars appearing as ``pink error models'' maintains presence but destroys identity and expression~\cite{VSM557272-cog-str}. Figure ~\ref{fig:pink-error-example} illustrates a common example of this phenomenon. These symptoms are particularly impactful in social XR applications where avatar appearance carries significant communicative and expressive weight.

\textbf{Incorrect Presence/Status Information (S3.3 Status Info)} manifested in 55/428 S3 bugs, such as Hubs-Foundation/hubs issue \#4528 where users appear ``In Lobby'' despite being present in the room. This is shown in Figure ~\ref{fig:Status-Info-example}, where a user's status is incorrectly displayed. While seemingly minor, these symptoms erode trust in the system's state representation and can lead to confusion about who is actually present and participating in the shared experience.

\subsubsection{Performance \& Stability Degradation (S4)}

Performance issues in multi-user XR carry heightened consequences compared to traditional applications, as they can induce physical discomfort and break presence.

\begin{figure}[t!] 
 \vspace{-0.5em} 
 	\centering
 	\subfigure[\parbox{0.8\linewidth}{Avatar Misrepresentation: Users' avatar appears as a generic "pink error model"}]{ 
 		\label{fig:pink-error-example} 
         \vspace{-1.5em} 
        \includegraphics[width=0.32\linewidth]{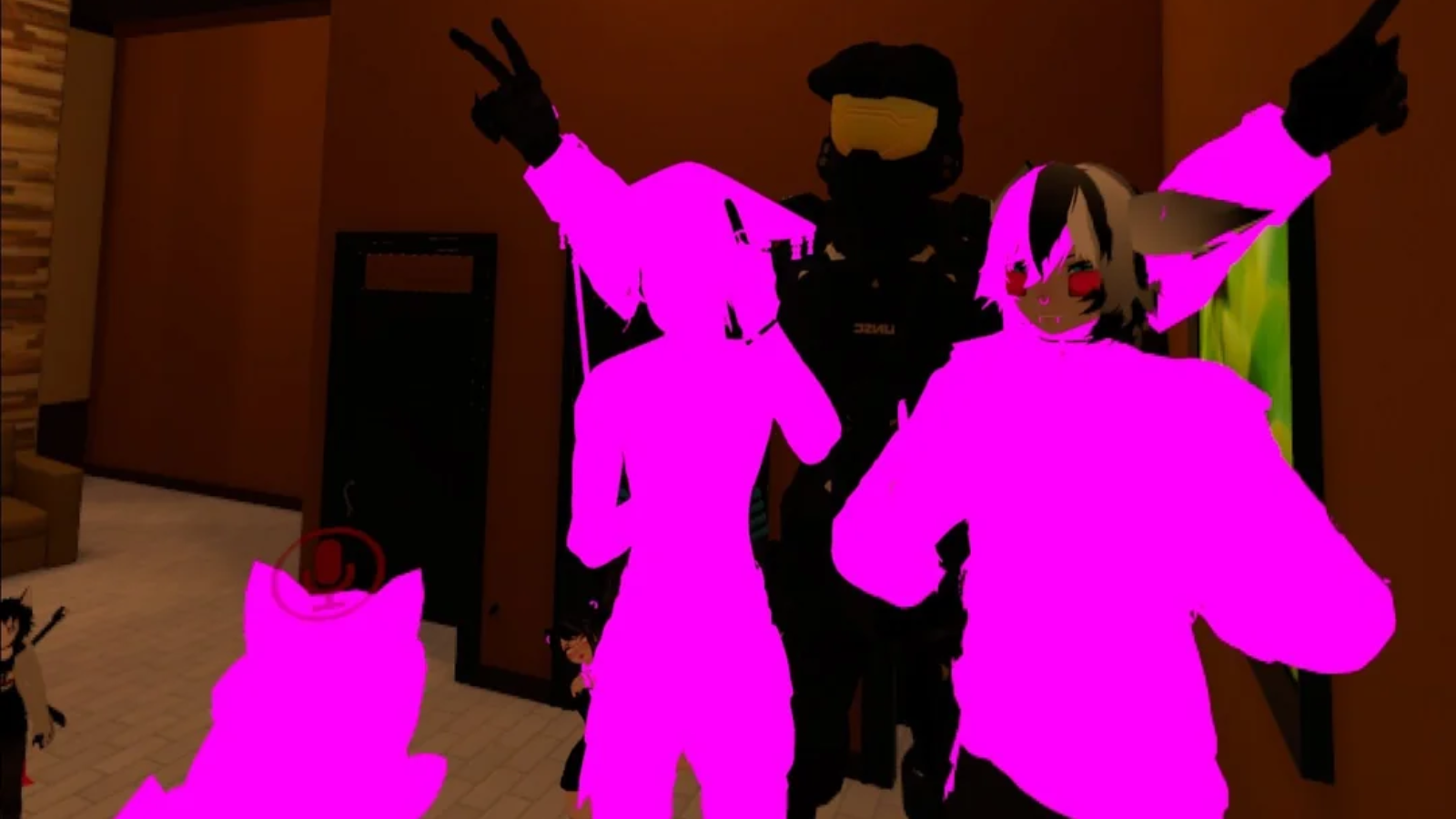}} 
 	\subfigure[\parbox{0.8\linewidth}{Screen Freezes: A frozen screen for an early joiner while a later joiner gets in without issue}]{ 
 		\label{fig:lag-example} 
 		\includegraphics[width=0.32\linewidth]{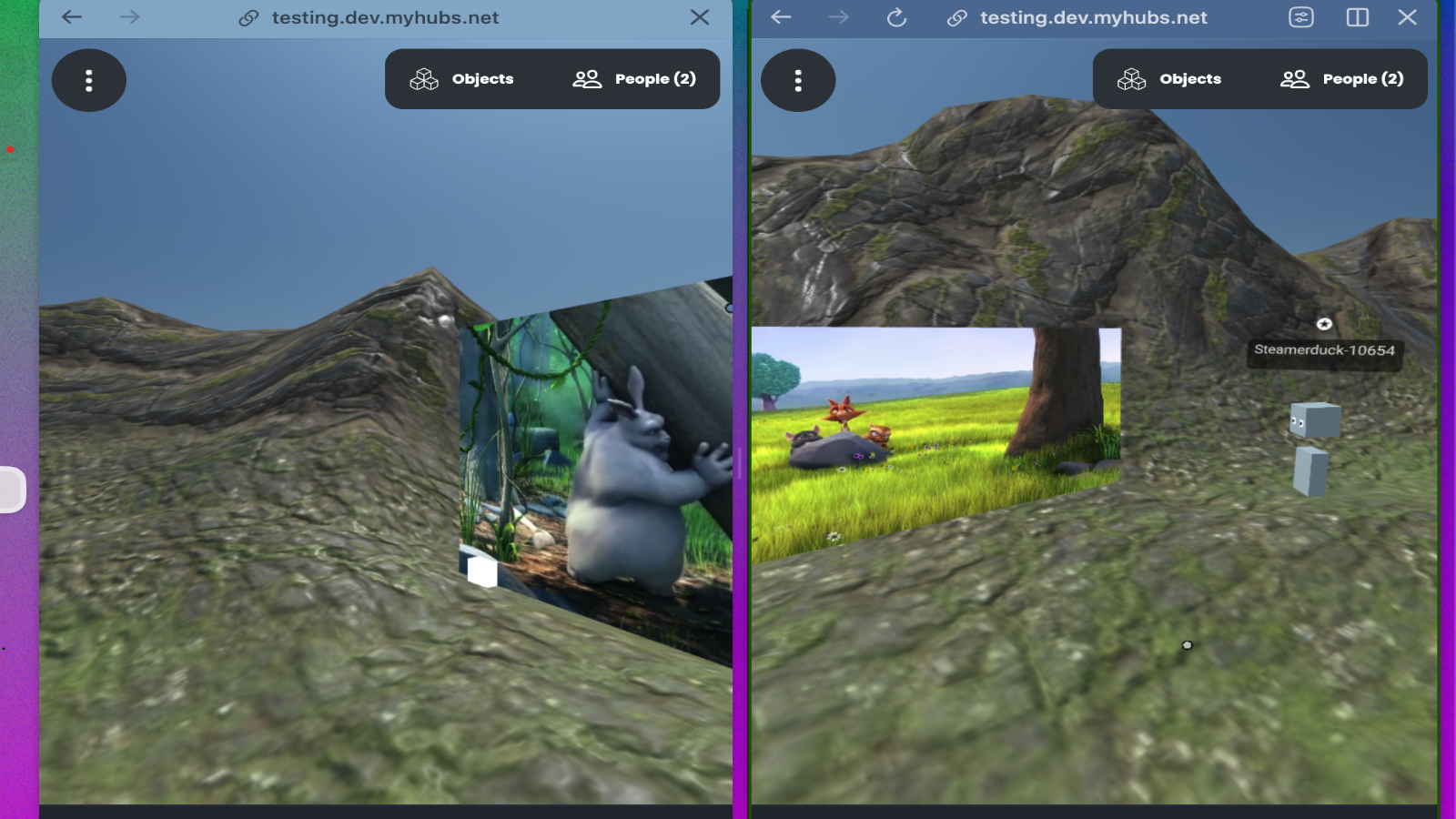}} 
 	\subfigure[\parbox{0.8\linewidth}{Disconnections: A user is unexpectedly kicked back to the game lobby}]{ 
 		\label{fig:connection-example} 
 		\includegraphics[width=0.32\linewidth]{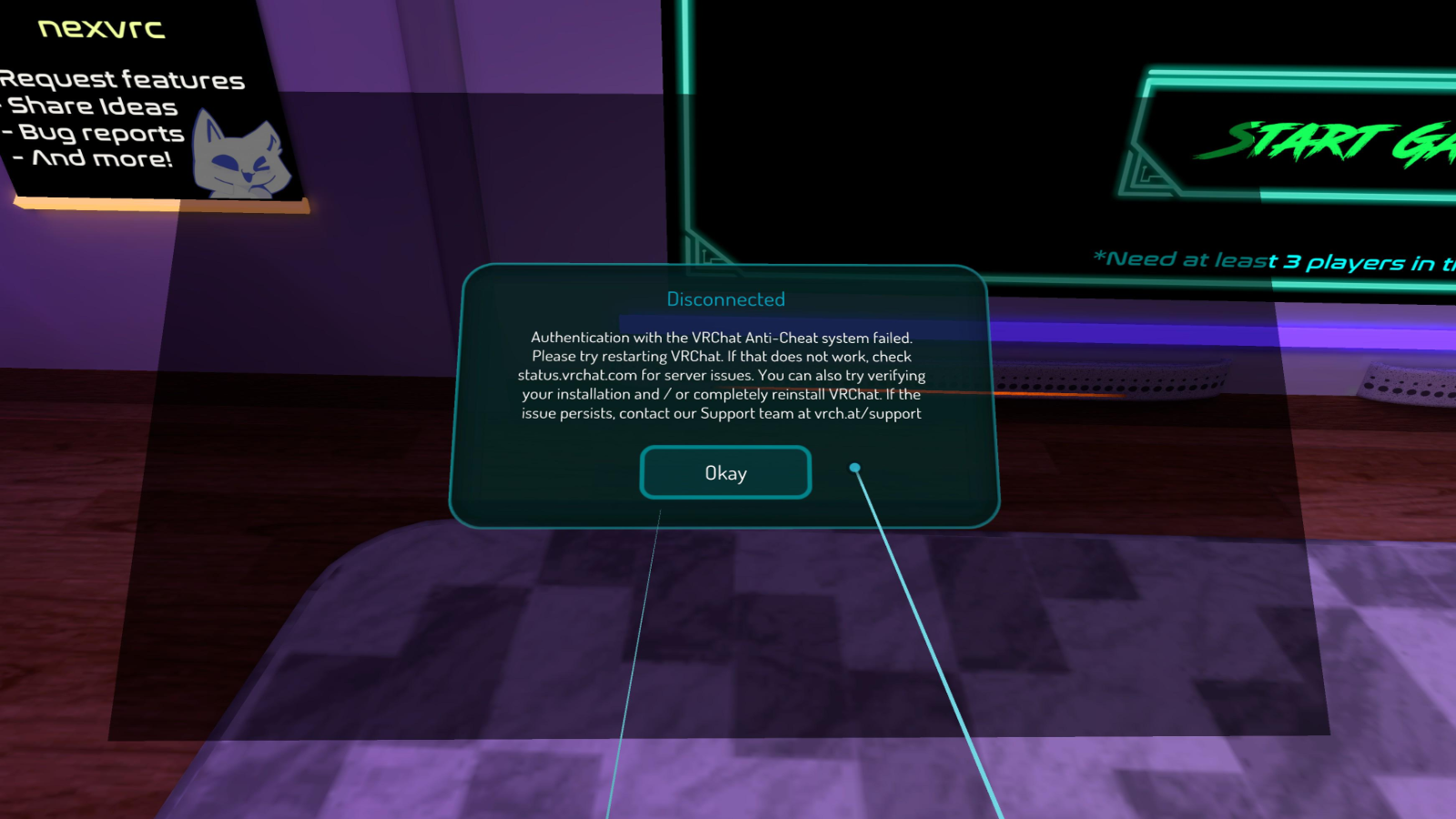}} 
      \vspace{-1.3em} 
       \caption{Example of Communication \& Awareness Impairment and Performance \& Stability Degradation} 
      \vspace{-1.5em} 
 	\label{figs:xr-device} 
 \end{figure}

\textbf{Connection Instability/Disconnections (S4.4 Connection)} represented the most severe symptoms (1,081/1,817 S4 bugs), with a notable one reporting disconnections ``every 3 minutes,'' making sustained collaborative sessions impossible~\cite{LHV688186-critical}. Another common issue is that users are mistakenly kicked back to the game lobby, preventing them from playing normally~\cite{OAA434570-disconnection}, as shown in Figure ~\ref{fig:connection-example}.

\textbf{Crashes/System Hangs (S4.3 System Crash)} affected 338/1,817 S4 bugs. A representative report of crashes ``30 seconds after entering a multiplayer server'' renders the application entirely unusable~\cite{NFE281823-critical}. The Hubs-Foundation/hubs issue \#6302, where ``non-privileged users attempting to grab objects crash the system,'' demonstrates how insufficient permission checking can escalate to complete system failure.

\textbf{Latency/Lag Spikes (S4.1)} and \textbf{Frame Drops/Freezes (S4.2) (Lag-Frame \& Freeze)} together accounted for 398/1,817 S4 bugs. A report description of ``massive lag spikes making all VR multiplayer virtually unusable'' illustrates how performance degradation can cascade across the entire experience~\cite{FFA309904-critical}. Another report of ``frequent frame skips causing tracking errors'' demonstrates the tight coupling between rendering performance and interaction accuracy in XR~\cite{HXR651273-major-func}. What is more serious is that, in the case of one Hubs, the person who joins later is able to join, but the person who joined earlier ends up with a frozen screen~\cite{UJF715696-freeze}, as shown in Figure ~\ref{fig:lag-example}.

\subsubsection{Cross-Cutting Patterns and Insights}

Several important patterns emerged from our symptom analysis:

Based on our analysis of bug reports from Developer Forums and User Communities—as the quality of user reviews was not sufficient for this purpose—several important patterns emerged from our symptom analysis:

\textbf{Temporal Dynamics}: A significant portion of symptoms (approximately 30.8\%) manifest specifically during session state changes—when users join, leave, or when ownership transfers occur. This suggests that transition states in multi-user XR systems are particularly fragile and deserve focused attention in both design and testing.

\textbf{Platform Heterogeneity}: Cross-platform scenarios frequently exacerbated symptoms, with 23.9\% of bugs mentioning platform-specific manifestations. This highlights the additional complexity introduced by %
diverse hardware ecosystem.

\textbf{Cascading Failures}: We observed that 6.1\% of bug reports described multiple co-occurring symptoms, suggesting that failures in multi-user XR systems often cascade. For instance, state desynchronization frequently led to interaction discrepancies, which in turn caused performance degradation.

\textbf{Immersion-Breaking Symptoms}: Unlike traditional software where bugs may be annoying but tolerable, 68.4\% of the identified symptoms directly break immersion or presence—core qualities of XR experiences. This elevates the severity of even seemingly minor symptoms.

The rich variety and high frequency of observable symptoms in multi-user XR systems highlight significant QA challenges. State synchronization issues are prevalent, suggesting that fundamental architectural decisions about distributed state management need careful reconsideration in XR contexts. Moreover, the tight coupling between different symptom categories indicates that holistic QA approaches may be more effective than targeting individual symptoms in isolation for improving multi-user reliability.

\subsection{RQ2: What are the underlying root causes contributing to these multi-user XR bugs?}

Understanding the root causes of multi-user XR bugs is essential for developing effective prevention and mitigation strategies. Through systematic analysis of the 106 bug reports, we identified and categorized the underlying technical origins of these defects. This analysis reveals that multi-user XR bugs stem from complex interactions between distributed systems, real-time synchronization requirements, platform-specific constraints, and application-level logic errors.

\subsubsection{Distribution of Root Cause Categories}

Table~\ref{tab:root-statistics} presents the distribution of root causes across the four main categories identified in our taxonomy. Network \& Synchronization Logic (RC1) emerged as the most prevalent category, accounting for 33.0\% (35/106) of the analyzed bugs. Session \& User Management (RC2) constituted 14.2\% (15/106), while XR Platform/SDK \& Hardware Integration (RC3) represented 27.6\% (29/106). Application Logic Errors (RC4) were the least common at 7.6\% (8/106).

\begin{table}[t!]
\centering
\caption{Root Causes Categories and Statistics\tnote{1}}
\label{tab:root-statistics}
\vspace{-1em}
\begin{threeparttable}
\resizebox{0.5\columnwidth}{!}{
\begin{tabular}{l|crr}
\hline
\textbf{Main Categories} & \textbf{Subcategories} & \textbf{Count} & \textbf{Percentage} \\
\hline
Net-Sync & Sync-Defect & 11 & 10.38\% \\
  & Race-Timing & 4 & 3.77\% \\
  & Net-Msg & 7 & 6.60\% \\
  & Net-Conn & 13 & 12.26\% \\
\hline
Session-User & Own-Flaw & 3 & 2.83\% \\
  & Join-Leave & 7 & 6.60\% \\
  & Perm-Ctrl  & 5 & 4.72\% \\
\hline
Platform-Hardware & SDK-Limit & 12 & 11.32\% \\
  & HW-Incompat & 8 & 7.55\% \\
  & Plat-Bug & 6 & 5.66\% \\
  & Res-Mgmt & 3 & 2.83\% \\
\hline
App-Logic & Inst-Remote & 3 & 2.83\% \\
  & Game-Rule & 5 & 4.72\% \\
\hline
Others & --- & 19 & 17.92\% \\
\hline
Total & --- & 106 & 100.00\% \\
\hline
\end{tabular}
}
\begin{tablenotes}
\footnotesize
\item[1] Some bug reports do not contain a verified fixing solution, thus hard to analyze root causes.
\end{tablenotes}
\end{threeparttable}
\vspace{-2em}
\end{table}

\subsubsection{Network \& Synchronization Logic (RC1 Net-Sync)}

The dominance of network and synchronization issues reflects the fundamental challenge of maintaining consistent shared state across distributed clients in real-time, immersive environments. Within this category, we identified four distinct sub-patterns:

\textbf{Defective State Synchronization Algorithms (RC1.1 Sync-Defect)} represented the most critical subcategory, accounting for 31.4\% (11/35) of RC1 bugs. These defects arise from fundamental flaws in how systems maintain consistency across clients. For instance, a bug in networked-aframe demonstrated a stateless replication approach that failed to handle entity deletions, when one client deleted a component, the deletion event was not properly propagated, leaving phantom entities on other clients~\cite{CEU902107-sync-algorithms}. Similarly, the untitled networked-aframe issue revealed that new clients joining a session received stale object positions because the synchronization logic failed to query and transmit the current authoritative state~\cite{KSI870368-sync-alg}.The fllowing Listing~\ref{lst:KSI} shows that all participants will fight for the ownership after connecting. the current owner of the present may not have the chance to give new joiner the current position.

\vspace{-1em}
\begin{lstlisting}[caption={Defective state synchronization algorithms example}, label={lst:KSI}]
onConnect() {
  setTimeout(() => {
    if (this.el.components.networked.data.owner === "scene") {
      console.log("Updating owner for:", this.el.id)
      NAF.utils.takeOwnership(this.el)
      this.update()
    }
  }, this.data.ownershipTimer)
\end{lstlisting}

The analysis revealed that many synchronization algorithms suffer from incomplete state representation. One of the typical examples is the issue where avatars would disappear because the \texttt{networkUpdate} method was invoked before component initialization completed, causing null reference exceptions that silently failed and prevented proper state replication~\cite{LXW738037-sync-algorithms}. This timing-dependent behavior suggests that many synchronization implementations lack robust initialization sequences and error handling.

\textbf{Race Conditions and Timing Issues (RC1.3 Race-Timing)} constituted 11.4\% (4/35) of RC1 bugs, highlighting the challenges of concurrent operations in distributed XR systems. These bugs manifest when the correctness of the system depends on the relative timing of events across clients. Hubs-Foundation/hubs issue \#4779 illustrated a classic ownership race condition: when multiple users simultaneously attempted to claim ownership of a media frame, the lack of atomic ownership transfer mechanisms led to split-brain scenarios where different clients believed they owned the same object.

Particularly problematic were race conditions during session state changes. One of bug reports demonstrated how broadcasting \texttt{syncAll} messages on every new data channel establishment caused entities to be instantiated multiple times when users joined rooms with existing participants~\cite{THB306317-race-timing}. The root cause traced to inadequate mutual exclusion mechanisms during the critical join synchronization phase.

\textbf{Flawed Network Message Handling (RC1.2 Net-Msg)} accounted for 20.0\% (7/35) of RC1 bugs. These defects encompass errors in message serialization, deserialization, ordering, and processing. For example, a bug report revealed how an error in the \texttt{sendData} method of a networking library could halt the entire render loop, as exception handling failed to isolate network errors from the main application thread~\cite{IWK186734-net-mes}. This coupling between networking and rendering layers violates fundamental architectural principles and leads to cascading failures.

Message ordering violations were particularly insidious. One of bug showed how incorrect cleanup of pending media requests led to screen sharing failures, messages for new sharing sessions were processed before cleanup messages from previous sessions completed, resulting in resource conflicts and silent failures~\cite{KTZ793597-net-mes}.

\textbf{Inadequate Network Connection Management (RC1.4 Net-Conn)} comprised 37.1\% (13/35) of RC1 bugs. These issues arose from insufficient handling of real-world network conditions and configurations. In the reports we have collected, there are three bugs exhibited the same pattern: one-way visibility or communication when users connected from different network segments. The root cause traced to missing TURN server configurations, preventing proper NAT traversal~\cite{BWY376541-net-con}~\cite{SDU920081-net-con}~\cite{YWP493007-net-con}. This pattern suggests that many XR applications are tested primarily in ideal network conditions and fail to account for the complexities of consumer network environments.

\subsubsection{Session \& User Management (RC2 Session-User)}

The second major category of root causes centered on the lifecycle management of users within shared sessions. These bugs often manifested during critical state transitions such as users joining or leaving sessions.

\vspace{-1em}
\begin{lstlisting}[caption={Flawed ownership management example}, label={lst:Own-Flaw}]
async _onPresenceUpdate(e) {
    try {
      let connectedIds = Object.keys(APP.hubChannel.presence.state || {});
      if (this.data.pinned) return;
      if (this.el?.components?.networked?.data?.persistent) return;
      let creator = this.el?.components?.networked?.data?.creator;
      let owner = this.el?.components?.networked?.data?.owner;
      if (!connectedIds.includes(creator) && !connectedIds.includes(owner)) {
        await NAF.utils.takeOwnership(this.el);
      }
    } catch (error) {
      console.warn(error);
    }
\end{lstlisting}

\textbf{Flawed Ownership Management (RC2.1 Own-Flaw)} accounted for 20.0\% (3/15) of R2 bugs. Object ownership, determining which client has authority over a shared object's state, emerged as a particularly error-prone concept. Hubs-Foundation/hubs issue \#5586 exemplified the complexity: when an object's creator left the session and another user interacted with it before also leaving, the object entered an orphaned state with no valid owner, preventing any future synchronization. Listing ~\ref{lst:Own-Flaw} code addresses this by checking if both the creator and the owner of an object have left the session. If both are gone, a user who is still present is assigned to take ownership, preventing the object from entering an unsynchronizable "orphaned" state.

The analysis showed that many systems implement ownership implicitly rather than as an explicit property. For example, in Hubs-Foundation/hubs issue \#1000, objects became invisible to new users if the last user to interact with them (the implicit owner) left the session. The fix—explicitly tracking and transferring ownership—indicates that robust multi-user systems need first-class ownership abstractions.

\textbf{Errors in Handling User Join/Leave (RC2.3 Join-Leave)} constituted 46.7\% (7/15) of RC2 bugs. These defects highlight the complexity of managing distributed state during dynamic membership changes. For instance, a bug revealed how inadequate synchronization during the join process caused entity duplication. When new users joined, the system broadcast full state snapshots to all participants rather than just to the joining user, causing existing users to re-instantiate already-present entities~\cite{THB306317-user-join}. A further example of a bug is that only the first person to join a room can be heard, while subsequent users cannot transmit audio\cite{RPQ265395-major-func}. The cause of this bug is that the code in Listing ~\ref{lst:Join-Leave} is called only once when the local stream is first created, rather than being automatically executed again when new users join. As a result, users who join later may not be added to the scope of this logic.

\vspace{-1em}
\begin{lstlisting}[caption={Errors in handling user join/leave example}, label={lst:Join-Leave}]
localStream.getTracks().forEach(
                track => {
                  Object.keys(self.peers).forEach(peerId => { 
                  self.peers[peerId].pc.addTrack(track, localStream) 
                })
              })
\end{lstlisting}

Leave-handling proved equally problematic. Hubs-Foundation/hubs issues \#4892 showed how pinned objects would disappear for all users when their creator left, even though the system design intended for pinned objects to persist. The root cause traced to cleanup logic that failed to distinguish between different object persistence classes.

\textbf{Incorrect Permission Control (RC2.2 Perm-Ctrl)} accounted for 33.3\% (5/15) of RC2 bugs. These security-critical defects arose from inadequate or incorrectly implemented access control mechanisms. Hubs issue \#6250 showed how the \texttt{hasPermissionToGrab()} function defaulted to \texttt{true} for entities missing specific components, allowing unauthorized object manipulation. This secure-by-default violation represents a common antipattern in distributed systems. Another classic case is when non-privileged users, with the "create and move objects" permission turned off, attempt to grab an object from a spawner, causing Hubs to crash~\cite{UCM387971-Perm-Ctrl}. This was resolved in Listing ~\ref{lst:Perm-Ctrl}, which adds a check for spawn permissions when grabbing the object spawner.

\vspace{-1em}
\begin{lstlisting}[caption={Incorrect permission control example}, label={lst:Perm-Ctrl}]
function* spawnObjectJob(world: HubsWorld, spawner: EntityID) {
+  if (!APP.hubChannel!.can("spawn_and_move_media")) return;
+
   const spawned = createNetworkedMedia(world, {
     src: APP.getString(ObjectSpawner.src[spawner])!,
     recenter: true,
     resize: true,
     animateLoad: true,
     isObjectMenuTarget: true
   });
\end{lstlisting}

\subsubsection{XR Platform/SDK \& Hardware Integration (RC3 Platfrom-Hardware)}

Platform and hardware-related issues formed the third major category, reflecting the heterogeneous ecosystem of XR technologies.

\textbf{SDK/API Misuse or Limitation (RC3.2 SDK-Limit)} represented 41.4\% (12/29) of RC3 bugs. These defects arose from incorrect usage of XR SDKs or encountering their inherent limitations in multi-user contexts. A bug illustrated a common Unity-specific issue: instantiating multiple OVRCameraRig components in a multiplayer scene caused camera control conflicts, as the SDK was designed assuming a single local player~\cite{DLU106275-SDK-API}. The Meta Community Forums revealed similar patterns with avatar SDKs, where lip sync functionality broke in multiplayer due to conflicts between Photon Voice's microphone usage and Meta Avatar SDK's LipSyncMicInput component~\cite{WIZ459768-SDK-API}.

\textbf{Hardware/Driver Incompatibility (RC3.3 HW-Incompat)} constituted 24.1\% (7/29) of RC3 bugs. These issues emerged from the interaction between software and specific hardware configurations. A Typical bug in Dying Light 2 traced to Proton version compatibility issues on Linux~\cite{JNY156322-hard-dri}, while another one revealed how background processes like ViveDashboard.exe could cause massive lag spikes in SteamVR applications~\cite{FFA309904-critical}. A particularly interesting case involved VRChat avatar rendering failures on specific AMD RX series graphics cards due to shader compilation bugs, demonstrating how low-level graphics driver issues can manifest as high-level multi-user problemst~\cite{TSU786578-AMD}.

\textbf{Platform/Engine-Specific Bugs (RC3.1 Plat-Bug)} accounted for 24.1\% (7/29) of RC3 bugs. A bug exemplified Unreal Engine-specific challenges: HMD tracking failed for clients in VR multiplayer projects due to incorrect assumptions in the engine's default VR template about authority and input replication. Apple's VisionOS SharePlay synchronization issues similarly revealed platform-specific challenges in adapting traditional multiplayer paradigms to spatial computing contextst~\cite{HHW685498-plat-eng}.

\textbf{Resource Management Issues (RC3.4 Res-Mgmt)} comprised 10.3\% (3/29) of RC3 bugs but had severe impacts. Hubs-Foundation/hubs issue \#5057 demonstrated how audio processing in rooms with 20-25 users caused complete audio failure on Android devices due to Panner Node performance limitations. The cascading nature of resource exhaustion, where CPU pressure from audio processing affected overall system stability, illustrates the importance of holistic resource management in multi-user XR.

\subsubsection{Application Logic Errors (RC4 App-Logic)}

While less frequent, application-specific logic errors revealed curcial patterns in how developers conceptualize multi-user interactions.

\textbf{Incorrect Local Instantiation/Handling of Remote Entities (RC4.2 Inst-Remote)} dominated at 37.5\% (3/8) of RC4 bugs. One of bugs provided a telling example: a user instantiating bullets couldn't see them because the client-side code used \texttt{document.createElement('a-entity')} instead of \texttt{document.createElement('a-sphere')}, omitting necessary geometry componentst~\cite{ADP259911-entities} .To address this issue, the template was modified to explicitly include an \texttt{<a-entity>} wrapper around the \texttt{<a-sphere>} element as shown in Listing~\ref{lst:bullets}. This error reveals a fundamental misunderstanding of how local and remote entity representations must maintain consistency while potentially differing in implementation details.

\vspace{-1em}
\begin{lstlisting}[caption={Incorrect instantiation example}, label={lst:bullets}]
   <template id="bullet-template">
-    <a-sphere class="bullet"
-      scale="0.1 0.1 0.1"
-      color="#fff"
-    ></a-sphere>
+    <a-entity>
+      <a-sphere class="bullet"
+        scale="0.1 0.1 0.1"
+        color="#fff"
+      ></a-sphere>
+    </a-entity>
   </template>
\end{lstlisting}

\textbf{Flawed Game/Interaction Rules for Multi-User (RC4.1 Game-Rule)} accounted for 62.5\% (5/8). A bug of AH-64 Apache multiplayer demonstrated how game-specific logic failed to account for multiple players: each player slotting into a new Apache created additional waypoint sets, causing navigation confusiont~\cite{QZS593885-rules}. These bugs suggest that developers often retrofit single-player logic for multiplayer scenarios without fully considering the implications of multiple actors.

\subsubsection{Cross-Cutting Observations}

Our analysis revealed several important cross-cutting patterns in root causes:

\textbf{Configuration Sensitivity:} A notable subset of bugs (approximately 14.98\%) were resolved through configuration adjustments rather than code changes ~\cite{TXW358683-config}~\cite{LHV688186-critical}. This pattern suggests that multi-user XR systems are highly sensitive to deployment configurations, and that many "bugs" may actually represent inadequate default configurations or poor configuration documentation.

\textbf{Layered Complexity:} Many bugs exhibited multiple contributing root causes across categories. For instance, avatar synchronization failures often involved both network synchronization logic flaws (RC1) and SDK integration issues (RC3). This layered complexity suggests that effective debugging requires understanding the full stack from application logic through networking to platform-specific implementations.

\textbf{Implicit Assumptions:} A recurring theme was the prevalence of implicit assumptions that hold in single-user contexts but fail in multi-user scenarios. Examples include assuming single camera rigs, single input sources, or that object creators persist for the object's lifetime. These assumptions, when not %
well
documented and handled, become fertile ground for bugs.

\textbf{Error Propagation:} The analysis revealed poor error isolation as a systemic issue. Network errors crashing render loops~\cite{KTZ793597-error-prop}, permission check failures crashing systems (Hubs \#6302), and resource exhaustion cascading to system-wide failures all demonstrate inadequate error boundaries between system components.

These findings suggest that addressing multi-user XR bugs requires not just fixing individual defects but rethinking fundamental architectural decisions around state management, error handling, and the explicit modeling of distributed system concepts like ownership and authority. The prevalence of synchronization and session management issues indicates that current frameworks and engines may provide insufficient abstractions for building robust multi-user XR experiences, leaving developers to repeatedly solve complex distributed systems problems in ad-hoc ways.

\subsection{RQ3: Typical Consequences on User Experience and System Functionality}

The analysis reveals that multi-user XR bugs lead to severe consequences, with 94.4\% resulting in critical failures (CS1) or major functionality impairments (CS2). Figure 4 presents the distribution of consequence severity.

\subsubsection{Critical System Failures (34.13\%)}

Critical failures manifest in three primary forms. \textit{System crashes} (e.g., Hubs \#6302) often stem from unhandled exceptions in permission systems, where unauthorized user actions trigger application-wide failures. \textit{Persistent disconnections} create particularly poor experiences through repeated connection failures every 3-5 minutes~\cite{LHV688186-critical}. \textit{Complete interaction breakdowns} ~\cite{FFA309904-critical} render multiplayer functionality "virtually unusable" through severe lag spikes, leaving users in applications that appear functional but lack core collaborative features.

\subsubsection{Major Functionality Impairments (60.32\%)}

Major impairments significantly degrade collaborative experiences without complete failure. \textit{Loss of sensory feedback} includes visual failures such as black cameras~\cite{MNX571081-major-func} and audio malfunctions (Hubs-Foundation/hubs \#5379: one-way communication), directly impacting spatial awareness and coordination. \textit{Perception misalignment} between users~\cite{WZE463732-avatar-state-desync} creates asymmetric shared realities that break interaction assumptions. \textit{Performance degradation} (eg., frame skips) causing tracking errors~\cite{HXR651273-major-func} exhibits non-linear scaling, functioning adequately with few users but degrading catastrophically beyond thresholds.

\subsubsection{Context-Dependent Amplification}

Bug consequences vary significantly by context. Educational applications show heightened sensitivity to synchronization errors due to potential learning of incorrect procedures. Session duration amplifies intermittent bugs exponentially—minor glitches become intolerable in extended work sessions. User expertise modulates impact: novices often blame themselves and abandon the technology, while experts develop workarounds but lose platform trust.

Table 3 summarizes the impact of different consequence types on key XR experience dimensions:

\begin{table}[t!]
\centering
\caption{Impact of Bug Consequences on XR Experience Dimensions}
\label{tab:consequence-impact}
\vspace{-1em}
\begin{threeparttable}
\resizebox{0.6\columnwidth}{!}{
\begin{tabular}{l|ccccc}
\hline
\textbf{Consequence} & \textbf{Presence} & \textbf{Immersion} & \textbf{Social} & \textbf{Task} & \textbf{Trust} \\
\hline
System Crash & *** & *** & *** & *** & *** \\
Disconnection & *** & *** & *** & *** & ** \\
Interaction Loss & ** & ** & *** & *** & ** \\
Sensory Loss & ** & *** & ** & ** & * \\
Perception Error & ** & ** & *** & ** & * \\
Performance & * & ** & * & ** & * \\
\hline
\end{tabular}
}
\begin{tablenotes}
\footnotesize
\item (*) Minor impact, (**) Major impact, (***) Severe impact

\end{tablenotes}
\end{threeparttable}
\end{table}

\subsection{RQ4: Privacy and Health Implications}

Our analysis identified 30 bugs with privacy implications and 45 bugs with health consequences, revealing unique risks in immersive multi-user environments.

\subsubsection{Privacy Vulnerabilities}

Privacy breaches occur through four primary vectors: 
(1) \textbf{Unintended Information Disclosure.} Audio bugs create severe risks when users unknowingly broadcast private conversations (such as first-user-only transmission~\cite{RPQ265395-major-func}) or experience one-way audio leaks. 
(2) \textbf{Behavioral Data Exposure.} Avatar synchronization bugs reveal sensitive information—incorrect height display~\cite{WZE463732-avatar-state-desync} exposes physical characteristics, while position history persistence reveals movement patterns and interaction preferences.
(3) \textbf{Unauthorized Access.} Control hijacking bugs like camera takeover~\cite{DLU106275-SDK-API} grant unintended access to other users' perspectives, potentially exposing confidential visual information in professional contexts.
(4) \textbf{Presence Artifacts.} Ghost avatars from disconnected users~\cite{OQG868838-pre-art} create false presence indicators, misleading others about occupancy in supposedly private spaces.

\subsubsection{Health and Safety Implications}

Multi-user bugs pose distinct health risks beyond traditional usability concerns:
(1) \textbf{Cybersickness Induction.} Performance bugs strongly correlate with vestibular discomfort. Frame skips~\cite{HXR651273-major-func} and unpredictable latency during user joins/leaves create visual-vestibular mismatches. Multi-user scenarios amplify these effects through compound synchronization delays.
(2) \textbf{Cognitive Stress.} Avatar invisibility~\cite{VSM557272-cog-str} creates "social presence uncertainty," leading to hypervigilance, communication anxiety, and decision paralysis in collaborative tasks. Users report increased cognitive load from compensating for unreliable social cues.
(3) \textbf{Physical Safety.} Control loss bugs~\cite{CTC541318-phy-safe} pose collision risks when users lose movement control during physical navigation. The sudden agency loss violates fundamental XR safety assumptions.
(4) \textbf{Cumulative Effects.} Prolonged exposure to intermittent bugs shows concerning health impacts. Persistent audio glitches (Hubs \#3927) cause headaches, tinnitus-like symptoms, and degraded spatial perception persisting post-session.

\subsubsection{Vulnerable Populations}

Certain users face disproportionate risks:
(1) \textbf{Vestibular-sensitive users} experience severe symptoms from performance bugs that cause mild discomfort in typical users.
(2) \textbf{Privacy-conscious professionals} in healthcare, legal, or corporate settings face elevated breach consequences.
(3) \textbf{Minors in educational settings} are vulnerable to both unauthorized behavioral data collection and prolonged performance degradation exposure.

\subsubsection{Design Implications}

Critical mitigation strategies emerging from our analysis include:
(1) \textbf{Privacy-Preserving Architectures.} Local-first processing of sensitive avatar data with selective synchronization of sanitized states. Granular user control over shared behavioral data with persistent transmission indicators.
(2) \textbf{Health-Aware Degradation.} Automatic complexity reduction during performance issues, disabling non-essential animations during synchronization problems, and clear warnings when health-impacting bugs are detected.
(3) \textbf{Safety Mechanisms.} Mandatory timeouts for unresolved health-impacting bugs, particularly those inducing cybersickness, preventing dangerous prolonged exposure.

The intersection of privacy and health concerns in multi-user XR reveals needs beyond traditional software quality metrics, establishing ethical imperatives for XR system design as adoption expands into sensitive applications.
\section{Discussion}

Our empirical study of 2,649 multi-user XR bugs reveals critical insights into the unique challenges facing this rapidly evolving domain. In this section, we discuss the implications of our findings for different stakeholders, reflect on the broader significance of our taxonomy, and discuss our limitations.

\subsection{Implications for Developers}

\subsubsection{Prioritizing Synchronization and State Management}
Our findings reveal that State \& Synchronization Inconsistency (S1) accounts for most representative category of symptoms, with 478 bug reports (18.04\%) exhibiting these issues. This dominance underscores a fundamental challenge in multi-user XR development: maintaining consistent shared state across heterogeneous devices and network conditions. Developers should invest substantial effort in designing robust synchronization protocols from the project's inception rather than treating them as implementation details to be addressed later.

The prevalence of Defective State Synchronization Algorithm/Logic (RC1.1) as a root cause suggests that many developers underestimate the complexity of distributed state management in XR contexts. Unlike traditional distributed systems, XR applications must synchronize not only discrete data but also continuous streams of spatial information, user movements, and complex 3D object states—all while maintaining the low latency required for immersion. We recommend adopting established patterns such as:
(1) \textbf{Authoritative server architecture} with client-side prediction and reconciliation, particularly for applications where consistency is paramount
(2) \textbf{Event sourcing} approaches that maintain a reliable history of state changes, enabling robust conflict resolution
(3) \textbf{Hierarchical state management} that distinguishes between critical shared state (e.g., game progression) and less critical local state (e.g., particle effects)

\subsubsection{Defensive Programming for Dynamic Sessions}
The high frequency of bugs related to user join/leave events (RC2.3, affecting 24 cases) indicates that session dynamics represent a particularly fragile aspect of multi-user XR systems. Developers must adopt defensive programming practices that assume users will join and leave at arbitrary times, potentially during critical state transitions. Key strategies include:
(1) Implementing comprehensive state snapshot mechanisms for late-joining users. 
(2) Ensuring graceful cleanup of user-owned resources upon disconnection. 
(3) Testing edge cases such as rapid join/leave sequences and simultaneous user actions. 

\subsubsection{Platform-Aware Development}
Our analysis reveals that Platform/Engine-Specific Bugs (RC3.1) and SDK/API Misuse (RC3.2) together account for 28 bug reports, highlighting the importance of deep platform knowledge. The complexity of XR development stacks—spanning game engines, XR SDKs, networking libraries, and hardware-specific APIs—creates many opportunities for integration failures. Developers should:
(1) Maintain comprehensive documentation of platform-specific behaviors and limitations. 
(2) Implement abstraction layers that isolate platform-specific code.
(3) Establish rigorous cross-platform testing protocols early.%

\subsection{Implications for Researchers}

\subsubsection{Need for Domain-Specific Software Analysis Techniques}
Traditional software analysis techniques often fail to capture the unique characteristics of multi-user XR bugs. For instance, static analysis tools designed for conventional distributed systems cannot reason about spatial relationships, embodied interactions, or the perceptual requirements of XR applications. Our taxonomy provides a foundation for developing specialized analysis techniques that consider:
(1) \textbf{Spatial consistency analysis}: Tools that can detect potential inconsistencies in how spatial data is synchronized across users.
(2) \textbf{Temporal analysis for XR}: Techniques that account for the strict latency requirements and frame-rate constraints of immersive applications.
(3) \textbf{Multi-modal interaction verification}: Methods to verify the consistency of audio, visual, and haptic feedback across distributed users.

\subsubsection{Empirical Studies on Bug Evolution}
While our study provides a snapshot of multi-user XR bugs, longitudinal studies are needed to understand how these bugs evolve as the technology matures.  
Research questions include:
(1) How do bug patterns change as developers gain experience with multi-user XR development?
(2) What is the relationship between platform evolution and bug introduction rates?
(3) How effective are different bug prevention strategies over time?

\subsubsection{Formal Methods for XR Systems}
The complexity of multi-user XR systems, combined with the severity of consequences we observed (with 20.65\% of bugs leading to Critical Failure or Major Functionality Impairment), suggests that formal verification techniques could provide significant value. Potential applications include:
(1) Model checking for ownership transfer protocols.
(2) Formal specification of synchronization invariants.
(3) Automated theorem proving for permission control systems.

\subsection{Implications for Tool Builders}

\subsubsection{Enhanced Debugging and Monitoring Tools}
Current debugging tools are ill-equipped to handle the distributed, real-time nature of multi-user XR applications. Our findings suggest several areas where specialized tools could dramatically improve developer productivity:
(1) \textbf{Distributed state visualizers} that can display the state of shared objects across all connected clients simultaneously.
(2) \textbf{Network replay tools} that can reproduce exact sequences of network messages to diagnose synchronization bugs.
(3) \textbf{Performance profilers} that correlate frame drops and latency spikes with specific multi-user events.

\subsubsection{Testing Frameworks for Multi-User Scenarios}
The diversity of symptoms and root causes in our taxonomy highlights the need for comprehensive testing frameworks specifically designed for multi-user XR. Such frameworks should support:
(1) Automated generation of multi-user test scenarios based on our symptom categories.
(2) Network condition simulation (latency, packet loss, jitter) integrated with XR testing.
(3) Regression testing that can detect subtle synchronization issues across builds.

\subsection{Implications for Platform and SDK Providers}

\subsubsection{Improved Default Behaviors and Documentation}
The prevalence of SDK/API Misuse (RC3.2) suggests that current XR platforms may not provide sufficiently intuitive or well-documented APIs for multi-user scenarios. According to bug report analysis, platform providers should consider:
(1) Establish secure and robust default configurations for common multi-user patterns.
(2) Provide comprehensive example implementations of multi-user features.
(3) Create diagnostic tools that can detect common misconfigurations.

\subsubsection{Built-in Support for Common Patterns}
Many of the bugs we identify stem from developers reimplementing common multi-user patterns incorrectly. Platform providers could significantly reduce bug rates by offering built-in support for:
(1) Ownership management with automatic conflict resolution.
(2) State synchronization with configurable consistency models.
(3) Session management with robust join/leave handling.

\subsection{Broader Implications}

\textit{The Unique Nature of Multi-User XR Bugs}
Our study reveals that multi-user XR bugs exhibit characteristics that distinguish them from bugs in traditional distributed systems or single-user XR applications. The tight coupling between physical and virtual spaces, the importance of embodied presence, and the need for precise spatial synchronization create a unique bug landscape that requires specialized attention from the software engineering community.

\textit{Economic and Social Impact}
With 2505 bugs (94.56\%) resulting in Critical Failure or Major Functionality Impairment, the economic impact of multi-user XR bugs is substantial. As XR technologies increasingly support critical applications in education, training, and remote collaboration, the cost of these failures extends beyond user frustration to include lost productivity, failed training exercises, and damaged user trust in XR technologies.

\subsection{Threats to Validity}

\textit{Internal Validity}.
Several factors may affect the internal validity:
In many cases, bug reporters describe symptoms without identifying the true root causes. Our classification of root causes sometimes relied on inference from available information, which may not always reflect the actual underlying issues.
Users may be more likely to report certain types of bugs (e.g., crashes) than others (e.g., minor synchronization glitches), 
which may skew our frequency analysis.

\textit{External Validity}.
The generalizability of our findings faces several limitations:
While we collected data from diverse sources, certain platforms or applications may be over- or under-represented in public bug reports.
The XR landscape evolves rapidly. Our findings represent a snapshot of bugs reported up to our data collection cutoff and may not fully reflect emerging technologies or practices.
Bug reports from different sources vary in technical detail and accuracy, potentially affecting our ability to correctly classify certain bugs.
\section{Related Work}
\label{sec:related-work}

\subsection{Exploratory Studies on XR Software}

Various works have conducted exploratory studies to understand XR software systems from different perspectives~\cite{li2020exploratory, li2023towards, DBLP:conf/esem/RodriguezW17, DBLP:conf/icse/NusratHZW21, DBLP:journals/tse/GuoDLXHZ25, li2024grounded, li2024xrzoo, li2025extended, li2024less}. 
Rodriguez and Wang~\cite{DBLP:conf/esem/RodriguezW17} analyzed the growing trends, project development structure, and challenges in VR software projects. 
Adams et al.~\cite{DBLP:conf/soups/AdamsBBMPR18} revealed the privacy and security threats in VR applications through interviews with users and developers. 
Miller et al.~\cite{miller2020personal} leveraged machine learning techniques to prove personal identifiability with user tracking data in VR software. Li et al.~\cite{li2020exploratory} modeled software quality of VR applications in 12 categories based on users' reviews. 
Nusrat et al.~\cite{DBLP:conf/icse/NusratHZW21} investigated performance optimization techniques for Unity-based VR projects and their impact on the software lifecycle. 
Abraham et al.~\cite{DBLP:conf/chi/AbrahamMK24} first integrated the immersive characteristics of AR with permission management, demonstrating the effectiveness of fine-grained control in enhancing privacy protection. 
Bose et al.~\cite{DBLP:conf/kbse/Bose024} conducted an empirical study on challenges faced by AR/VR developers. 
Rzig et al.~\cite{paper:vr-testing-empirical} performed a large-scale empirical study on software testing practices of open-source VR projects.
Guo et al.~\cite{DBLP:journals/tse/GuoDLXHZ25}  evaluated the security vulnerabilities and privacy data leaks of VR applications. 
Gu et al.~\cite{gu2025software} first conducted a systematic mapping study on software testing for XR applications.

\subsection{Studies on Multi-user XR Software System}

Several works have proposed empirical studies or approaches to understand or address challenges in multi-user XR software systems. Xanthidou et al.~\cite{xanthidou2024collaboration} conducted a survey and recognized spatial design, collaborative interactions between users and VR environments, and video/audio fidelity as major challenges in VR collaboration. 
Ruth et al.~\cite{DBLP:conf/uss/RuthKR19} explored the challenges in designing secure and private content sharing for multi-user AR software.
Tümler et al. ~\cite{DBLP:conf/hci/TumlerTY22} evaluated the experience in multi-user multi-platform XR systems. 
Zhang et al.~\cite{DBLP:journals/ase/ZhangRGLL25} systematically studied the dilemma between immersive user experience and higher privacy risks in different usage scenarios of the Metaverse. 
Rasch et al.~\cite{DBLP:conf/chi/RaschRS023} compared different intention communication methods for locomotion in multi-user VR scenes. 
Merz et al.~\cite{DBLP:conf/vr/MerzGWL24} investigated the influence of different IO device characteristics and degrees of immersion on the user experience in asymmetric virtual collaboration. 
Dong et al.~\cite{DBLP:conf/vr/DongSGF21} proposed a dynamic density-based walking redirecting method to reduce collision in multi-user virtual environments. 
Chen and Guo~\cite{DBLP:journals/network/ChenG23} designed a task offloading scheme to improve the quality of service in multi-user mobile AR services.

\section{Conclusion}

Multi-user Extended Reality (XR) systems pose distinct SE challenges due to complex real-time synchronization and immersive interactions. Through empirical analysis of 2,649 real-world bug reports, we develop comprehensive taxonomies categorizing symptoms, root causes, and consequences of XR defects.
Synchronization inconsistencies, avatar anomalies, and performance issues were prominent, driven mainly by network logic flaws and session management errors. Notably, over 34\% of defects cause severe disruptions like crashes and interaction breakdowns.
Our findings underline unique privacy and health risks in multi-user XR contexts, recommending robust synchronization practices, defensive session handling, and enhanced platform documentation. 
\balance
\bibliographystyle{ACM-Reference-Format}
\bibliography{multiuser}

\end{document}